\documentclass[sigconf]{acmart}

\setcopyright{none}
\renewcommand\footnotetextcopyrightpermission[1]{}
\usepackage[acronym,nowarn]{glossaries}
\usepackage[]{todonotes}
\usepackage{url}
\usepackage{marvosym}
\usepackage{ragged2e}
\usepackage{array}
\usepackage{booktabs}
\usepackage{subcaption}
\usepackage{newfloat}
\usepackage{footmisc}
\usepackage{multirow}
\usepackage{xspace}
\usepackage{multicol}
\usepackage[T2A,T1]{fontenc}
\usepackage[utf8x]{inputenc}
\usepackage{enumitem}
\usepackage{soul}

\graphicspath{{figures/}}

\newcommand{\dns}[1]{{\small \texttt{#1}}}

\newcommand{\rf}{\dns{\fontencoding{T2A}\selectfont .\cyrr\cyrf}\xspace}
\makeatletter
\DeclareRobustCommand\onedot{\futurelet\@let@token\@onedot}
\def\@onedot{\ifx\@let@token.\else.\null\fi\xspace}

\def\eg{\emph{e.g}\onedot} 
\def\ie{\emph{i.e}\onedot}

\def\etal{\emph{et al}\onedot}
\makeatother



\author{Raffaele Sommese}
\affiliation{%
  \institution{University of Twente} 
  \country{}
}
\email{r.sommese@utwente.nl}

\author{Roland van Rijswijk-Deij}
\affiliation{%
  \institution{University of Twente} 
  \country{}
}
\email{r.m.vanrijswijk@utwente.nl}

\author{Mattijs Jonker}
\affiliation{%
  \institution{University of Twente} 
  \country{}
}
\email{m.jonker@utwente.nl}

\renewcommand{\shortauthors}{R. Sommese, R. van Rijswijk-Deij, and M. Jonker}

\begin{document}

    \fancypagestyle{firstpagestyle}{%
        \fancyhf[H]{} %
	\fancyfoot[L]{\small \emph{arXiv pre-print}}%
	\fancyfoot[R]{\small \emph{Under peer review}}%
    }

    \fancypagestyle{plain}{%
        \fancyhf[HL]{} %
        \fancyhf[HR]{\shortauthors} %
	\fancyfoot[L]{\small \emph{arXiv pre-print}}%
	\fancyfoot[R]{\small \emph{Under peer review}}%
    }


\begin{abstract}

Domain lists are a key ingredient for representative censuses of the Web.
Unfortunately, such censuses typically lack a view on domains under
country-code top-level domains (ccTLDs). This introduces unwanted bias:
many countries have a rich local Web that remains hidden if their
ccTLDs are not considered. The reason ccTLDs are rarely considered is
that gaining access -- if possible at all -- is often laborious. To
tackle this, we ask: what can we learn about ccTLDs from public
sources? We extract domain names under ccTLDs from 6 years of public
data from Certificate Transparency logs and Common Crawl. We compare
this against ground truth for 19 ccTLDs for which we have the full DNS
zone. We find that public data covers 43\%-80\% of these ccTLDs, and
that coverage grows over time. By also comparing port scan data we then
show that these public sources reveal a significant part of the Web
presence under a ccTLD.  We conclude that in the absence of full access
to ccTLDs, domain names learned from public sources can be a good proxy
when performing Web censuses.

\end{abstract}


\title[]{This Is a Local Domain: On Amassing Country-Code Top-Level Domains from Public Data}

\maketitle

    \pagestyle{plain}


\section{Introduction}
\label{sec:introduction}

The Web is studied for a multitude of reasons. A common starting point for such
studies is a list of domain names. These can be so-called top lists of popular
domains~\cite{Pochat2019}, or lists of domain names extracted from the DNS zone
files for top-level domains (TLDs). The latter are a particularly valuable
source, as TLD zone files give a comprehensive view on a well-defined part of
the name space. For so-called \textit{generic} TLDs (gTLDs), such as the
well-known triad of \texttt{.com}, \texttt{.net} and \texttt{.org}, but also
more recent additions such as \texttt{.berlin} or \texttt{.shop}, these zone
files can be obtained through a process defined by ICANN. By agreeing to a set
of guidelines on not disclosing significant parts of these zone files to the
public, users of ICANN's Centralized Zone Data Service can download these zone
files and use them for various purposes including research.

The Web, however, extends well beyond the generic TLDs. Coun\-try-code
top-level domains (ccTLDs) are TLDs specific to a country, and often represent
a large part of the local Web~\cite{Baeza2007}. Unfortunately, gaining access
to domain lists for ccTLDs is nowhere near as easy as gaining access to gTLDs.
If access is possible at all, the process is often laborious and requires
lengthy personal interactions with the governing body of the ccTLD and the
signing of restrictive contracts. As a consequence, domains in ccTLDs are
understudied, and studies that do exist are often incomplete, limiting
themselves to a handful of ccTLDs researchers have gained access to. Some
researchers resort to the use of commercially available domain lists, but the
provenance and reliability of these lists is unknown. The root of this problem
is that ccTLD governance is not under ICANN's purview, but instead a matter of
local policy~\cite{Merrill2016}. Yet studying ``\textit{the} Web'' while
omitting ccTLDs leads to unwanted bias, as Web domains under ccTLDs often
reflect a rich local or regional ecosystem~\cite{Baeza2007, Zembruski2021}. In
fact, historians argue that the Web is now so ingrained in our culture, that it
must be archived for posterity. Yet they run into the same challenges as
network and security researchers if they want to archive national Web
collections~\cite{Bruegger2019}.

To address this issue, we turn to alternative sources: public datasets that
contain significant numbers of domain names. Specifically, we look at two
sources: Certificate Transparency (CT) logs~\cite{rfc9162} and Common
Crawl~\cite{CommonCrawl} data. Based on longitudinal datasets covering the six
years from 2017 to 2023, we extract second-level domains (SLDs) registered
under 19 ccTLDs. We then compare this to ground truth data (full domain lists)
which we obtain from the OpenINTEL project~\cite{openintel}, to establish to
what extent and how timely public data sources cover the full list of names in
these ccTLDs. Then, we take a recent snapshot and use port scan data to look at
the extent to which data from public sources covers domains with hosts that
have port 80 (HTTP) and 443 (HTTPS) open compared to the full list of domains
in the ccTLD. This leads to the following contributions:

\noindent
\begin{itemize}[leftmargin=*]
	\item We show that public data sources can cover 43\%-80\% of all names
		in 19 ccTLDs, and show that coverage grows over time.
	\item We break down what the individual public data sources contribute
		and show that while CT logs offer a solid basis, adding Common
		Crawl data can increase coverage by up to 11 percentage points.
	\item Using port scan data, we demonstrate that domain lists from
		public sources provide substantial coverage of domains with an
		active Web presence.
	\item We analyse the timeliness of public data sources and show that
		60\% of newly registered domains appear in CT logs within a day
		of registration, and 80\% show up within five days.
	\item As there are over 300 ccTLDs in existence, we validate our
		coverage against gTLDs (for which we also have full domain
		lists) and find a comparable coverage range.
\end{itemize}

Our results show that in the absence of access to full domain lists for ccTLDs,
data from public sources may serve as a good proxy for active Web domains. More
importantly, coverage is steadily growing, in step with the increasing use of
TLS~\cite{Holz2020}, which in turn requires domain owners to request
certificates that thanks to browser enforcement are now almost
universally recorded in public CT logs. This paves the way for future studies
that encompass the Web in full, including a local and regional view.


\section{Methodology and Data}
\label{sec:data_and_methodology}

In this section we introduce the data sources from which we amass domain names as well as
the ground truth dataset that we use specifically for this paper. We then detail
our approach to using these datasets to study coverage.

\subsection{Amassing Domain Names}
\label{sec:data_and_methodology:mining}

We amass domain names from two publicly available and diverse data sources: \emph{CT logs}
and \emph{Common Crawl}. We selected these particular sources because of their
sustained availability and scale.

\vspace{0.5em}
\noindent
{\bf CT Logs --}
Certificate Transparency (CT) is a protocol that enables the logging of TLS
certificates for monitoring and auditing purposes~\cite{rfc9162}. Using CT,
certificate authorities (CAs) record issued certificates in so-called
certificate logs. These public, append-only logs can be monitored by virtually
anyone, so as to audit CA activity and to pick up on the issuance of suspect
certificates.
Various parties operate CT logs, including CAs such as Let's Encrypt and
DigiCert, and browser vendors such as Google. Leading browser
vendors also require TLS certificates to appear in a log. Since 2018, both
Apple and Google require newly issued certificates to be logged in CT logs.
Given that the combined market share of Apple and Google (including Chromium-based
browsers) is approaching 95\%,\footnote{\href{https://gs.statcounter.com/browser-market-share}{https://gs.statcounter.com/browser-market-share}} CAs are effectively forced
to log certificates intended for use in HTTPS in multiple CT logs (both
Apple~\cite{AppleCT} and Google's~\cite{GoogleCT} policy require recording in multiple logs).
Interestingly, the CA/Browser (CAB) forum has -- to date -- not adopted CT 
as a mandatory practice.

We established a pipeline to scrape CT logs and continually collect and store TLS certificates from CT logs in active
operation. At present, this dataset covers a total of 38 different CT logs
and many billions of unique certificates contained collectively therein.
The \texttt{Common Name} and \texttt{Subject Alternative Name} attributes of
X.509 certificates (can) contain domain names. By surveying these attributes, we
can garner scores of domain names from CT logs.

\vspace{0.5em}
\noindent
{\bf Common Crawl --}
Common Crawl is a nonprofit organization that builds and
maintains a sizable, open repository of Web crawl data, offering years and
petabytes of Web page data. The Common Crawl data lives in Amazon S3 as part of
Amazon's Open Data Sponsorship Program and is free for anyone to access.
Crawls are seeded from a set of candidate domain names and the crawler
follows links leading to other pages. Crawls are performed approximately every
one to two months and contain raw Web page data, metadata and text extractions,
among others. Relevant to our work, crawls accumulate many tens
of millions of registered domain names that one can extract from the
so-called URL index.

\subsection{Consolidated Dataset of Names}
\label{sec:data_and_methodology:dataset}

For this paper, we create a consolidated dataset using domain names amassed from CT log and
Common Crawl data. Next to their sustained availability and scale, we also
consider that these two sources will complement one another. The reason we consider
them as complementary is that they collect their source materials through
independent, disjoint processes. We hypothesize that this makes it likely that
the number of distinct domains in a union of both sources is larger than in each
data source individually.

From CT log data we consider all logged certificates with a validity starting
from 2017.  This selection spans 5.6 billion unique certificates, from which we
extract 390 million unique registered domain names (i.e., \dns{TLD+1}) in
total.
For Common Crawl we consider data for crawl snapshots dated between June~2017
and June~2023 (inclusive). There are 58 such snapshots, collectively accounting
for 127 million registered domain names.
The combined total number of unique registered domain names in our consolidated dataset is 430 million.

\subsection{Ground truth from ccTLD Zones}
\label{sec:data_and_methodology:baseline}

We need a baseline of ccTLD zones to investigate the extent to which domain
names amassed from CT log and Common Crawl data cover ccTLD zones. As ground truth, we leverage data
from the OpenINTEL project~\cite{openintel}. OpenINTEL is a large-scale, active DNS
measurement platform. Its (forward) DNS measurement is primarily seeded using zone files,
which are obtained under sharing agreements with registries. This measurement covers around 65\% of the
global DNS namespace. Notably and relevant to our study, OpenINTEL covers 19 of
the over 300 ccTLDs in existence. The ccTLDs that OpenINTEL covers are:
\dns{.at}, \dns{.ca}, \dns{.ch}, \dns{.co}, \dns{.dk}, \dns{.ee}, \dns{.fi}, \dns{.fr},
\dns{.gt}, \dns{.li}, \dns{.na}, \dns{.nl}, \dns{.nu}, \dns{.ru}, \dns{.se},
\dns{.sk}, \dns{.su}, \dns{.us}, and \rf. Two of these ccTLDs are
among the global Top 10 rank of ccTLDs (\dns{.nl} and \dns{.ru})~\cite{cctldstat}.
Our ground truth dataset spans from 2018 until 2023 and covers 51 million ccTLD
domain names in total.

\begin{figure}[t]
    \centering
    \includegraphics[width=0.9\columnwidth]{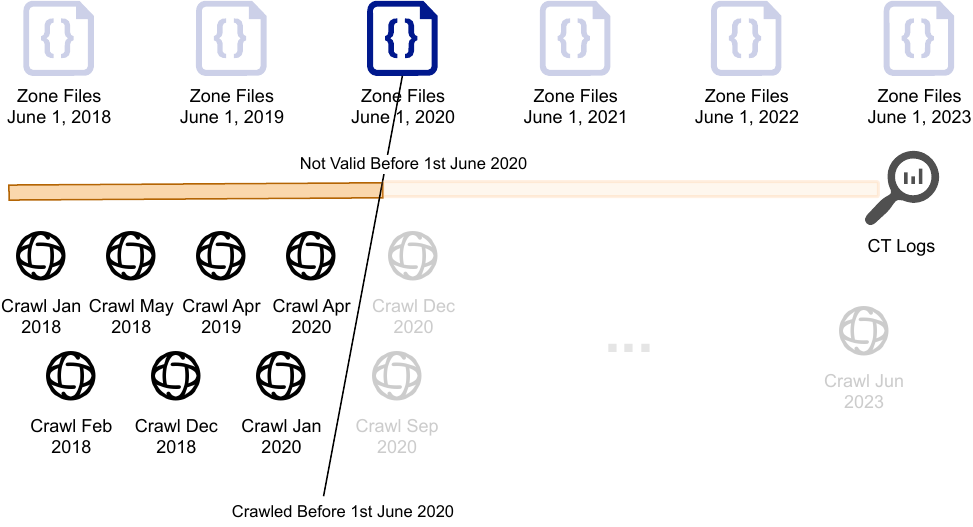}
    \caption{Example representation of our methodology for analyzing coverage considering the 1\textsuperscript{st} of June 2020.}
    \label{fig:methodology}
\end{figure}

\subsection{Quantifying Coverage}
\label{sec:data_and_methodology:methodology}

Our methodology for analyzing coverage is as follows.
We consider the consolidated dataset of ccTLD domain names from CT logs and Common Crawl data (see
Section~\ref{sec:data_and_methodology:dataset}), and our ground truth dataset based on ccTLD zones from
OpenINTEL (see Section~\ref{sec:data_and_methodology:baseline}). Our consolidated
dataset contains, in addition to amassed domain names, various metadata. In
particular, we also extract, where applicable, certificate validity start and
end timestamps, the names of CT logs in which certificates were recorded, and crawl
IDs and dates.
To conduct our longitudinal analysis, we cross-reference names amassed from the
public data sources against ground truth on a yearly basis. We choose a
yearly frequency based on the fact that domain names normally have a lifespan
of at least one year~\cite{Affinito2022}. We pick the first of June as cut-off
point for our analysis every year.
For the yearly baseline comparison, we consider from the consolidated
dataset all ccTLD domains with a certificate validity start date prior to the
analysis date across all CT logs, as well as all domains crawled by Common Crawl before June 1\textsuperscript{st}. \autoref{fig:methodology} illustrates the example of coverage analysis for June 1\textsuperscript{st},
2020. From the CT logs we consider any domain name from a certificate with a validity date prior to
June 1\textsuperscript{st} and from Common Crawl any name we learned from crawl data
preceding June 1\textsuperscript{st}, 2020. From the ground truth dataset, we consider all ccTLD domains
that are \textit{actually active} (\ie, delegated in the zone) on June 1\textsuperscript{st}.

We use the result from cross-referencing the consolidated dataset with the
ground truth dataset to study coverage and perform various longitudinal analyses.
We also use port discovery data to infer if domains have a Web purpose, by
looking for open HTTP and HTTPS ports. To this end, we use a smaller, auxiliary dataset
of ports scans of the IPv4 address space. These scans are run weekly from the University of
Twente network under IRB approval (ethics reference number 220032).


\section{Related Work}
\label{sec:related_work}

The most closely related work to ours is that of Hageman
\etal~\cite{hageman21}.  In their paper, they developed a tool called
\emph{Gollector}, which extracts domain names from various data sources,
namely CT logs, gTLD zone files, and passive DNS (collected at both recursive
resolvers and TLD authoritatives). They quantify how many names can be amassed
from various vantage points over a three-week period and make inter-vantage
point comparisons. The authors do not study the degree of coverage nor do they
compare against a baseline. The authors acknowledge the potential of employing
CT logs as an early detection mechanism, as we also discuss in Section~\ref{sec:results:lag}.

VanderSloot \etal~\cite{VanderSloot2016} studied the TLS certificate ecosystem.
Certificate scans are central to their work and they use various sources of
domain names to maximize certificate discovery. Specifically, they considered
data from CT logs, Censys, Common Crawl, IPv4 address space scans, zone files, and Top lists.
Their results indicate that CT log data covers 90.5\% of discoverable certificates,
with an additional 8.9\% coverage provided by Censys data. In our study, we focus on
domain names, and hence place an emphasis on coverage notably different from their
work.
Scheitle \etal focused on certificate transparency in particular and conducted a
large-scale analysis of CT practices by comparing data obtained from CT logs
with actively scanned domain names~\cite{gasser2018}.

Other works in the literature have a more privacy-oriented focus. Roberts
\etal~\cite{roberts2019} demonstrated how domain names extracted from CT logs
and Censys scans expose personal names, usernames, and email addresses.
Finally, Scheitle~\etal illustrated the leakage of DNS information from CT logs,
demonstrating how Fully Qualified Domain Names (FQDNs) are exposed in the logs~\cite{scheitle2018}.

In summary, works exist in the literature that involve domain extraction from
public sources in one form or another. To the best of our knowledge, ours is the first to
thoroughly investigate coverage by public sources and to compare against a ground truth.


\section{Results}
\label{sec:results}

\begin{table*}[tb]
\centering
\begin{tabular}{l||rr|r|rrr|rr}
\toprule
& \multicolumn{8}{c}{Number of ccTLD domain names and degrees of coverage (\% Total)} \\
    Date & {Total} & {Covered} & {Not Covered} & {$\in CT \smallsetminus CC$} &
    {$\in CC \smallsetminus CT$} & {$\in CC \cap CT$} & {$\in CT$} & {$\in CC$} \\
\midrule
2023-06-01 & 32.9\,M & 19.5\,M (59\%) & 13.4\,M (41\%) & 9.3\,M (28\%) & 2.4\,M~$$~$$~(7\%) & 7.9\,M (24\%) & 17.2\,M (52\%) & 10.3\,M (31\%) \\ \hline
2022-06-01 & 28.1\,M & 16.0\,M (57\%) & 12.1\,M (43\%) & 7.3\,M (26\%) & 2.4\,M~$$~$$~(9\%) & 6.3\,M (22\%) & 13.6\,M (49\%) &  8.7\,M (31\%) \\ \hline
2021-06-01 & 26.6\,M & 14.6\,M (55\%) & 11.9\,M (45\%) & 6.3\,M (24\%) & 2.7\,M (10\%) & 5.6\,M (21\%) & 12.0\,M (45\%) &  8.3\,M (31\%) \\ \hline
2020-06-01 & 25.2\,M & 13.0\,M (51\%) & 12.2\,M (49\%) & 5.1\,M (20\%) & 3.1\,M (12\%) & 4.7\,M (19\%) &  9.8\,M (39\%) &  7.9\,M (31\%) \\ \hline
2019-06-01 & 20.6\,M &  9.6\,M (46\%) & 11.1\,M (54\%) & 3.3\,M (16\%) & 3.2\,M (16\%) & 3.1\,M (15\%) &  6.4\,M (31\%) &  6.3\,M (31\%) \\ \hline
2018-06-01 & 20.9\,M &  7.7\,M (37\%) & 13.3\,M (63\%) & 2.1\,M (10\%) & 3.8\,M (18\%) & 1.7\,M~$$~$$~(8\%) &  3.9\,M (19\%) &  5.5\,M (26\%) \\
\bottomrule
\end{tabular}
\caption{CT logs (\emph{CT}) and Common Crawl (\emph{CC}) coverage in the yearly assessments.
    \emph{Total} names from the baseline ccTLD zone data (OpenINTEL) and \emph{Covered}
    by the consolidated dataset of domain names from public sources (CT and CC data).}
\label{tab:splitting}
\end{table*}

\subsection{Longitudinal Coverage}
\label{sec:results:longitudinal}

\begin{figure}[t]
    \centering
    \includegraphics[width=0.9\columnwidth]{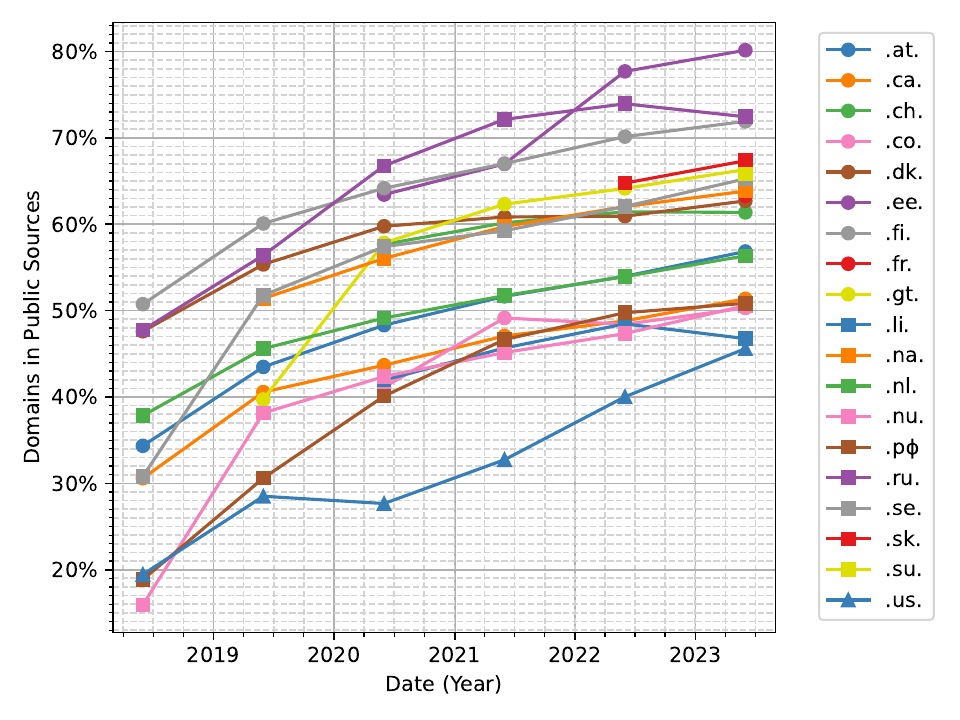}
    \caption{Extent of ccTLD zone coverage over time by the two public data sources
    under consideration (\ie, CT logs and Common Crawl).}
    \label{fig:public_sources}
\end{figure}

Our initial analysis is concerned with the development of coverage over the
five-year period of our study (2018 --- 2023). In \autoref{fig:public_sources},
we present the outcome of this analysis.
We draw two implications from our analysis here. First, we observe that in the
year 2023, the percentage of ccTLD zones covered by the dataset of names amassed
from CT logs and Common Crawl data varies significantly between ccTLDs. Specifically, the
coverage spans from a lower bound of 43\% for \dns{.us} domains, to a
substantial upper bound of 80\% for \dns{.ee} domains. The calculated weighted
average over the total set of domains yields a coverage of 59\%, which approximates the degree of coverage for a
substantial portion of the ccTLD zone files within a few percentage points.

The degree of coverage for \dns{.ee} stands out. This is, however, not that 
helpful, as \dns{.ee} is one of the few ccTLDs that publicly releases
their zone file. Much more captivating instances
are illustrated by ccTLDs that do not publicly share their zone
files. Specifically, when examining the Top 5 zones in
\autoref{fig:public_sources}, \dns{.ru} (Russian Federation) and \dns{.su} (Soviet Union) stand out.
These are challenging zones to access due to the ongoing conflict and
geopolitical events, yet 73\% of \dns{.ru} and 66\% of \dns{.su} can be learned
from CT and Common Crawl data. Also, \dns{.fi} as one of the Top 5 covered
ccTLDs, is another zone that can only be obtained under contract. This clearly
illustrates the added value of amassing ccTLD domains from public data sources.

Our second observation is the substantial increase in coverage over the past
years, as evidenced by \autoref{fig:public_sources}.  Back in 2018, the average
coverage was at 37\%, with a minimum of 15\% for \dns{.nu} and a maximum of 50\%
for \dns{.fi}.  In a time span of five years, this coverage has risen by 22
percentage points to the 59\% average in 2023, signifying an impressive
1.6$\times$ increase from the initial value.
We speculate that the reasons behind this substantial increase can primarily be
attributed to the widespread adoption of TLS, the democratization of the TLS
landscape (\ie, through free TLS certificates) and the concomitant hard
requirement from leading browser vendors to record certificates in CT logs.

\subsection{Coverage Contributions}
\label{sec:results:cc_vs_ct}

Having presented the overall longitudinal view, we now delve deeper into
coverage and dissect the contributions of the two public data sources that we
leverage in this study. We aim to understand their individual impacts on
enhancing the visibility of ccTLD zones.

In \autoref{tab:splitting}, we present the breakdown of contributions
from both CT logs and Common Crawl to the coverage of ccTLD zone files.
In the year 2023, CT logs account for the majority of the coverage.
Out of 19.5\,M domain names amassed from CT logs and 
Common Crawl data, 7.9\,M names appear in both sources,
while 9.3\,M are solely covered by CT logs, and 2.4\,M 
are exclusively covered by Common Crawl. In percentages, 24 points of the 59\%
average coverage (see Section~\ref{sec:results:longitudinal}) are due to
co-appearing names, 28 points are thanks solely to CT, and 7 points are
exclusively due to CC.
These results show that, although the vast majority of domain names
can be amassed from CT logs alone, there is additional value
in extracting domain names from other data such as Web crawl data.
Another interesting result arising from our analysis is that the supplemental
coverage provided by Common Crawl in comparison to CT logs is gradually
decreasing over time, even though the number of ccTLD domain names in Common
Crawl data is not itself decreasing.  This phenomenon might be linked again to
the continued expansion of TLS adoption, resulting in more ccTLD domain names becoming
discoverable via CT.

Turning our attention to CT logs, we now focus on the extent to
which a single log can offer sufficient coverage for ccTLDs domain names,
to aid choosing which logs to prioritize during obtainment efforts.
Our analysis reveals that when considering individual logs (recall from
Section~\ref{sec:data_and_methodology:mining} that our CT data encompasses 38
logs), the maximum achievable coverage hovers around 84--89\% of the total
number of domains covered by CT logs. This finding remains consistent across the
entire five-year period.
For the past three years, the logs that provided the highest coverage were
instances of the temporally-sharded Google \texttt{Xenon} log, specific to the
year in question (\ie, \texttt{Xenon-2021} provides the best coverage in the
June~2021 baseline comparison).\footnote{Temporally sharded logs contain TLS
certificates with a validity that ends in the sharded period. Many logs are
sharded for scaling purposes. Sharding is often done by year, but nowadays even by first
and second half of a year.}
\texttt{Xenon} is followed in the ranks by Google \texttt{Argon}, Cloudflare
\texttt{Nimbus}, and Let's Encrypt \texttt{Oak}. All of these are temporally
sharded. This underscores that the logs associated with the
present year best cover domain names in ccTLD zones.
In and before 2021, non-temporally-sharded Google logs such as
\texttt{Rocketeer} and \texttt{Pilot} led the individual coverage. Google
retired its last non-sharded logs in May of 2022.

Finally, we examined the contribution of domain names amassed specifically from
expired certificates. Our analysis reveals that approximately 19\% of the
domains acquired from CT logs in 2023 were amassed from expired certificates.
This percentage has exhibited a gradual increase over the years. This finding
underlines that expired certificates should not be ignored while amassing
domain names.
Altogether, our findings demonstrate the importance of considering multiple logs
in their entirety, preferably all available logs, to achieve the best coverage.

\subsection{Web Presence}
\label{sec:results:web}

\begin{table}[b!]
\begin{tabular}{lrr}
\toprule
IPv4 Address Record & \multicolumn{1}{c}{No} & \multicolumn{1}{c}{Yes} \\
\midrule
Both & 3.5\% & 96.5\% \\
CT logs & 7.7\% & 92.3\% \\
Common Crawl & 6.8\% & 93.2\% \\
Neither & 22.4\% & 77.6\% \\
\bottomrule
\end{tabular}
\caption{IPv4 address record presence per category. Percentages of domains in the
    respective set.}
\label{tab:addresses}
\end{table}

Arguably, the leading reason for domain name registration is to facilitate Web
content hosting. Similarly, the need to secure Web network traffic between a
site and its visitors is among the leading reasons to obtain a TLS certificate.
As previously explained in, Section~\ref{sec:data_and_methodology:mining},
browser enforcement of CT log inclusion has been a major driver for the adoption
of CT.
Consequently, we investigated the extent to which domain names that appear in CT
relate to the full set of domain names under a ccTLD zone seemingly
registered for Web-related purposes.
The first criterion we used for identifying potential Web-use purpose involves
checking for the presence of IPv4 address records in the DNS (\ie, \dns{A}
records).  Although such a record can signal domain name use for various other
services such as FTP, SSH, etc., it serves as a method for initially filtering
domains that are not resolvable and hence not likely to be in use for Web
purposes.
As shown in \autoref{tab:addresses}, the vast majority (96.5\%) of domain names
that appear both in CT logs and Common Crawl have IPv4 address records in the
DNS. Following are domains exclusively present in Common Crawl (93.2\%) and CT
Logs (92.3\%). On the contrary, among domains absent from both Common Crawl and
CT logs, only 77.6\% feature IPv4 address records. The remaining 22.4\%
potentially indicate domains that are not currently in active use or have been
registered for other purposes such as mail-only domains.
These results show that, generally, domain names learnable from CT log and Common Crawl
data are likely to have an \dns{A} record associated to them.

We now turn our attention to active services hosted on these IPv4 addresses.
More specifically, we look for the presence of open Web ports using the data
detailed in Section~\ref{sec:data_and_methodology:methodology}.
The results of the analysis, shown in
\autoref{tab:ports}, follow a pattern similar to the \dns{A} record presence.
Specifically, the majority of domain names learned from public sources offer service on a Web port. In
contrast, the percentage drops to 70.5\% for ccTLD domain names that are neither
learnable from CT logs nor from Common Crawl data. Names learned from the public sources under
consideration thus show a tendency to be used for Web purposes. This, of course,
one can partially expect, as Common Crawl's focus lies with Web
crawling, and browser enforcement should lead to CT inclusion.
On this note, an interesting aspect is that domains found in both CT
logs and Common Crawl, or exclusively in CT logs, have higher rates of
HTTPS deployment (87.7\% and 81.4\%, respectively), compared to those learnable
solely from Common Crawl, and significantly surpassing those in neither of the public sources
under consideration (50.6\%).
While the reason behind this result may be straightforward (\ie, certificates
used for HTTPS hosting are recorded in CT logs), this effectively demonstrates that 
leveraging CT logs data gives researchers a substantial coverage of the part of
the ccTLD zone used for secure Web deployment.

\begin{table}[t]
\begin{tabular}{l|cc|ccc}
\toprule
Web Ports & No & Yes & HTTP Only & HTTPS Only & Both \\
\midrule
Both & ~$$~~8.5\% & 91.5\% & ~$$~~3.1\% & 0.8\% & 87.7\% \\
CT & 13.1\% & 86.9\% & ~$$~~5.0\% & 0.6\% & 81.4\% \\
CC & 17.4\% & 82.6\% & 17.9\% & 0.4\% & 64.2\% \\
Neither & 29.5\% & 70.5\% & 19.7\% & 0.2\% & 50.6\%\\
\bottomrule
\end{tabular}
\caption{HTTP(s) presence per category. Percentages of domains in the respective set}
\label{tab:ports}
\end{table}

In a small percentage of cases in \autoref{tab:addresses} and
\autoref{tab:ports}, for Common Crawl we saw a surprising lack of \dns{A}
records in OpenINTEL measurements and no open HTTP(S) port. There are a couple
of reasons for this seeming inconsistency. 
First, it is important to note that OpenINTEL measurements and Common Crawl scans happen
at different times and in different places. These differences can affect the
results we see. Secondly, the way in which we amass domains from Common Crawl data
involves considering all names found in crawls leading up to a specific date,
not just in the most recent crawl. This means we may include domains that were
used for Web purposes in the past but are no longer being used for that purpose anymore. In
this case, the domains could lack \dns{A} records in later DNS measurements.

\subsection{Delay in Publication}
\label{sec:results:lag}

\begin{figure}[t]
    \centering
    \includegraphics[width=0.9\columnwidth]{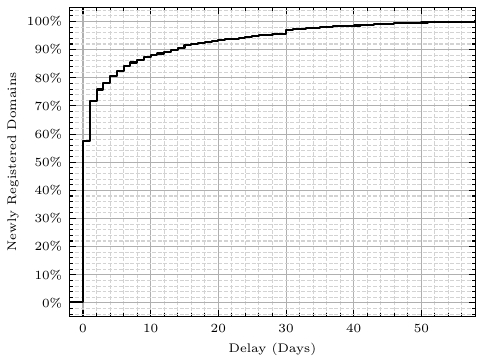}
    \caption{Lag between name publication in \dns{.sk} ccTLD zone file and
    appearance in CT logs.}
    \label{fig:lag}
\end{figure}

In addition to studying the degree of ccTLD zone coverage, we investigated
the extent to which public sources lag behind zone
publications. In simpler terms, we aim to understand how long it takes for a
ccTLD domain name to appear in public CT logs or Common Crawl data. While a possibly steep delay is
evident for Common Crawl due to its measurement frequency, the question is
more relevant for CT logs owing to their continuous operation.

Scraping CT logs does not provide an explicit and reliable indicator of the
publication time of a certificate. This is due to the way certificates are
appended to logs; there is a delay between CA issuance and log appending.
However, in a separate experiment unrelated to this paper, we collected, from a
real-time feed of certificates just recorded in CT, a dataset of
certificates for the \dns{.sk} ccTLD. This auxiliary dataset covers a
three-month period (March~2023 --- May~2023).
Leveraging OpenINTEL measurement data, we extract
recently registered domains within the full \dns{.sk} zone. By
comparing these two datasets, we sought to analyze the time it takes for a
domain to appear in CT logs after its creation.
Our findings are summarized in Figure~\ref{fig:lag}. The CDF illustrates that
nearly 60\% of newly registered \dns{.sk} domains appear in CT logs on the same day
they appear in the zone, with 80\% appearing within five days.
Although due to its limited scope, this result is not universally applicable
per se, it does suggest that CT logs could serve as a viable resource for obtaining
information about newly registered domain names reasonably close to their registration dates. In specific cases 
(\ie, when zone publishing is delayed), CT logs could even offer insights ahead of time.
This could prove remarkably significant, for example, to shed light on the infrastructure
of short-lived newly registered domain names, which are often exploited for malicious purposes~\cite{Affinito2022}.


\section{Can We Generalize These Results?}
\label{sec:gtlds}

\begin{figure}[t]
    \centering
    \includegraphics[width=0.9\columnwidth]{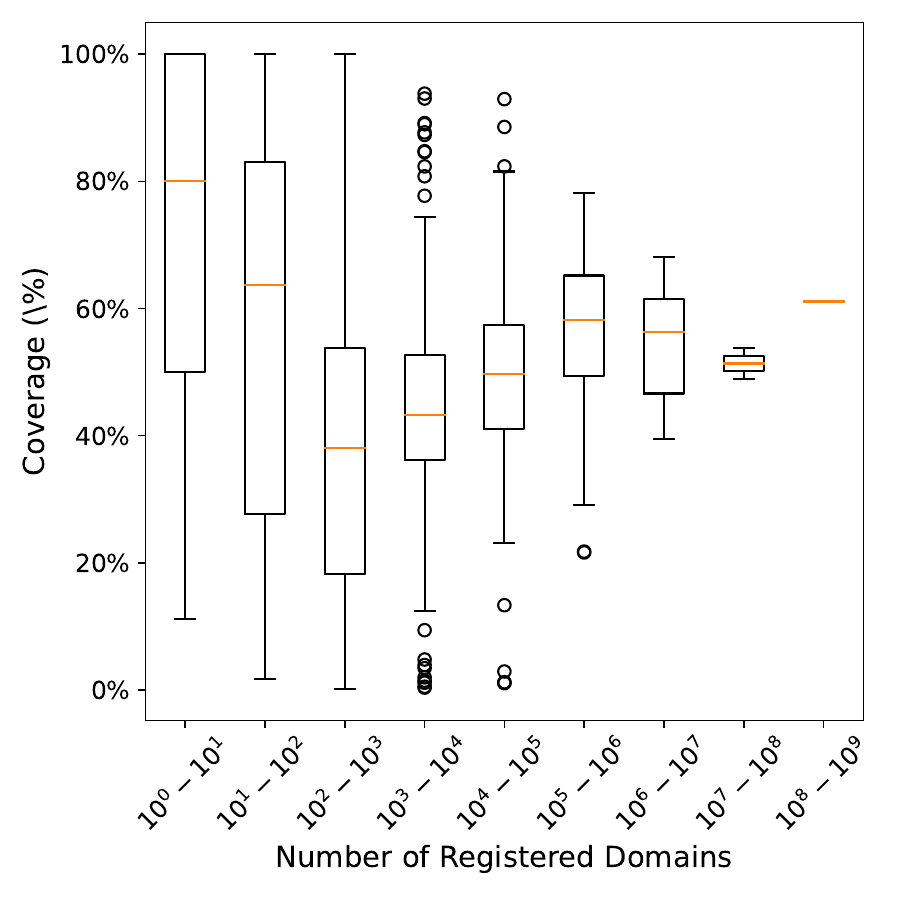}
    \caption{Coverage in 2023 of public sources over 1153 gTLDs.}
    \label{fig:gtlds}
\end{figure}

We previously demonstrated that the coverage of the analyzed ccTLDs spans from
43\% to 80\% in 2023 (see Section~\ref{sec:results:longitudinal}).  While this
outcome is promising, it is important to note that our ground truth data for 19 ccTLD zones
is small compared to the over 300 ccTLDs in existence and may be somewhat
biased towards specific regions of the world (e.g., we have a higher
representation of European ccTLDs).
To strengthen the foundations of our analysis, we chose to extend the coverage
assessment to gTLDs as well. Domain names from gTLD zones are obtainable via the
ICANN CZDS program and are longitudinally available in OpenINTEL data as well.

Considering the substantial number of available gTLDs (1153 as of 2023), we
decided to categorize them based on their volume of names. The
results for 2023 are shown in \autoref{fig:gtlds}.  Starting with
the largest gTLD (\ie, \dns{.com}), we observe that the coverage remains
around 60\% for this zone. Slightly lower figures are shown for gTLDs containing
between 10 and 100 million names (e.g., \dns{.net} and \dns{.org}).
The most interesting result, however, is that the average coverage across all
magnitudes falls within the range of 38\% to 80\%, with larger TLDs generally
displaying higher coverage rates -- excluding a few outliers. For the category of
smallest gTLDs (1--100 domains), several instances of high coverage are evident,
sometimes reaching up to 100\%.  We primarly identified this phenomenon in gTLDs
utilized for trademarks by companies (such as \dns{.android},
\dns{.oracle} \& \dns{.ferrero}).
While these results may not be universally applicable in cases where the
population of ccTLDs significantly diverges from that of gTLDs (\ie, domain registration
allowed only for specific business types, mandatory physical presence, or other 
special requirements), they do suggest
that our considerations regarding ccTLD coverage can generally be extended to
other ccTLDs. Overall, our findings suggest that the public sources under
consideration provide substantial coverage of ccTLD zones.


\section{Discussion}
\label{sec:discussion}

We discuss our work from several angles, ranging from the data used and
other sources, to registry considerations for sharing data.

\vspace{0.5em}
\noindent
{\bf Other public data sources --}
We focused in particular on CT logs and Common Crawl data because of their sustained
availability and scale (see Section~\ref{sec:data_and_methodology:mining}). There are other public Web crawl data sources that one could
consider. One example is the HTTP Archive, which is seeded with popular
domain names and has a smaller scale~\cite{http_archive}.
On the certificate side of things, one could consider certificates collected
from TLS scans of the IPv4 address space in addition to CT log data.

\vspace{0.5em}
\noindent
{\bf Implications of amassing from CT logs and Common Crawl --}
Given the nature of Web crawl and CT log data, and as confirmed by our results, one is
likely to find domain names used for Web purposes in these sources. Whether CAs
record certificates issued for other purposes in CT logs is beyond the scope of this
paper, but we expect TLS scans for the (default) certificate on non Web ports (\eg, on \texttt{465/TCP} for
SMTPS) to lead to the discovery of additional ccTLD names used for other purposes.

\vspace{0.5em}
\noindent
{\bf Feasibility of amassing domain names --}
We recognize that it requires resources to scrape and analyze CT log
data at scale. CT logs are retired and taken offline over time, but as our results show
the current (and thus active) CT logs provide the best coverage for the current
state of ccTLD zones.
On the Common Crawl end, domain names are readily available from its so-called URL index,
and crawl data is historically available.

\vspace{0.5em}
\noindent
{\bf Our appeal to registries --}
As our results show, sizable portions of ccTLD zones are already public. In our
opinion, this should detract from privacy concerns. Given the large community
benefits, registries could follow the lead of those that have already made their
zone public (\eg, IIS for \dns{.nu} and \dns{.se} and SWITCH for \dns{.ch} and
\dns{.li}).
We are aware that some registries have concerns related to matters other than
privacy for keeping their zones closed (\eg, commercial incentives). We believe
these concerns could be addressed by publishing the zone with a delay of a few
hours or days.


\section{Conclusions}
\label{sec:conclusions}

In this paper, we studied the degree of ccTLD coverage provided by domain names
amassed from public data. We amassed names from two public sources -- Common Crawl and
Certificate Transparency logs -- because of their sustained availability and scale,
and compared against ground truth for 19 ccTLD zones.
Our findings reveal that coverage has increased over the past years, with an
average of 59\% of ccTLD zones covered by the considered public data. We attribute this to the
uptake of TLS and Certificate Transparency.
Using auxiliary service port discovery data, we show that the public data gives
researchers a substantial coverage of names used for Web purposes. We also
investigated the delay with which names appeared in public data and reasoned
about applicability of our findings to ccTLD zones for which we do not have
ground truth.

\vspace{0.5em}
\noindent
{\bf Future Work --}
Recognizing that it requires resources to amass domain names in a manner similar
to ours, we plan to release ccTLD domain names to the community in the future.  This
release is planned as part of an expansion of the daily OpenINTEL measurement, which we will
seed with public data so we can also make the resulting measurement data public.

    \section*{Acknowledgements}
We want to thank Moritz M\"{u}ller for his help as we were shaping our paper,
and Gustavo L. Cesar and Ralph Holz for their efforts towards instrumenting the
campaign that provided supplemental port scan data. 
This work was partially funded by the Netherlands Organization for Scientific
Research (NWO) under grant number NWA.1215.18.003 (CATRIN).
This research was made possible by OpenINTEL, a joint project of the University
of Twente, SURF, SIDN, and NLnet Labs.

    \onecolumn
    \begin{multicols}{2}
        \bibliographystyle{ACM-Reference-Format}
        \bibliography{main}
    \end{multicols}
\end{document}